\documentstyle[12pt,epsfig]{article}
\def\hp{H^+}
\def\mh{m_{\hp}}
\def\tv{\tau^+\nu}
\def\cs{c\bar{s}}
\def\tb{\tan\beta}
\def\ctb{\cot\beta}
\begin{document}
\vspace{1.cm}
\begin{center}
\Large\bf QCD and SUSY-QCD corrections to the Three-Body Decay of the Charged Higgs Boson
\end {center}
\vspace{0.5cm}
\begin{center}
{ Xiao-Jun Bi, Yuan-Ben Dai and Xiao-Yuan Qi}\\\vspace{3mm}
{\it Institute of Theoretical Physics,
 Academia Sinica, \\ P.O.Box 2735, Beijing 100080, P. R. China }
\end{center}
\vspace{1.5cm}
\begin{abstract}
The ${\cal O}(\alpha_s)$ QCD corrections to the three-body decay width of the charged Higgs $\Gamma$($H^+\rightarrow$$W^+b\bar{b}$) are discussed in the MSSM model. Our calculations indicate that the standard QCD corrections to the three-body decay mode raise the width by about 12$\%$ and the supersymmetric QCD corrections(due to $\tilde{g}, \tilde{t}, \tilde{b}$ exchanges) can be comparable to or even larger than the standard QCD corrctions in some regions of the supersymmetric parameter space. This is mainly due to the effect of large left-right mixing of stop($\tilde{t}$). It could significantly affect the phenomenology of the $H^+$ search.
\end {abstract} 
\section{Introduction}
\paragraph{}
Despite of the great success of the standard model (SM) in elementary particle physics, one important aspect, the Higgs mechanism, of the model has not yet been experimentally verified, and there is plenty of room for extensions. The SM possesses the minimal Higgs structure of one doublet and only one neutral physical Higgs boson is left after SSB. One of the most interesting version of the extended SM is the minimal supersymmetric standard model(MSSM) which demands two Higgs doublets and predicts a charged Higgs pair($H^\pm$)\cite{mssm}. Search for charged Higgs will give bounds on the parameter space of new physics models. Several groups have analyzed their experimental data and give lower bounds on the charged Higgs mass
\cite{delph,cdf}.

The top quark decay provides a promising laboratory to look for Higgs boson because the top quark--Higgs Yukawa coupling is large. The $H^\pm$ search strategy in top-quark decays has been based on the distinctive features of the channels $t\rightarrow bH^+ \rightarrow b(\tau^+\nu,c\bar{s})$, compared with the standard model decay $t\rightarrow bW^+\rightarrow b( \ell\nu,\tau\nu,q' \bar{q})$, which assumes that the dominant decay channels of the charged Higgs below $t\bar{b}$ threshold are 
$\hp\rightarrow\tv$ and $\hp\rightarrow\cs$\cite{2}. Recently, another important three-body decay channel of the Higgs boson $\hp\rightarrow\bar{b}bW^+$ has been proposed by Ernest Ma, {\sl et al.}\cite{ma}. They found that the above three-body channel is dominant for $\hp\geq 140GeV$ and $\tb\leq 1$ while the $\tv$ mode dominates at large $\tb$. The $c\bar{s}$ mode is relatively small at all $\tb$ for $\hp\geq 140GeV$. This can significantly affect the signature for top quark products.

Accurate predictions of charged Higgs width including ${\cal O}(\alpha_s)$ corrections to the above processes are important to the phenomenology of the $\hp$ search. The one-loop effects (strong and electroweak) on the decay $t\rightarrow\hp b$ and $t\rightarrow bW^+$ have been discussed in detail in \cite{sola} and \cite{tbw} respectively. The leading QCD corrections to the process $\hp\rightarrow\cs$ is taken into account by substituting the quark mass parameters by the running masses at the $\hp$ mass scale\cite{n} which changes the charm quark mass from 1.5GeV to 1GeV and considerably reduces the width of the $\hp\rightarrow\cs$ process. Correspondingly, the leading QCD corrections to the above three-body decay channel should also be taken into account, which, so far as we know, is absent in the literatures. In this paper, we present the calculations of the QCD and SUSY-QCD corrections to the width of the three-body decay $\hp\rightarrow\bar{b}bW^+$ within the MSSM. We also calculate the one-loop corrections from charged Higgs particle which may be important for small $\tb$.

The paper is organized as follows. In section II, we present most of the formulae in our calculations and show some important analytical results. We also discuss the arrangement of the ultraviolet and infrared divergences. Some relevant formulae are given in Appendix A and B. In section III, numerical results and discussions are presented.

\section{Analysis and Formulae}
\paragraph{}

We shall perform the calculations mainly in the MSSM. The related pieces of Lagrangian are given in Appendix A. For simplicity, we neglect terms directly proportional to $m_b$ but of course keep those terms which are singular in the vanishing mass limit($\sim$ $\ln(m_b)$). In the final results after we include real gluon emission, such singularities are cancelled out. As a result of taking $m_b=0$ in the MSSM, only left-handed bottom-squark($\tilde{b}_L$) enters our calculations while the left-right handed mixing of the top-squark ($\tilde{t}_L, \tilde{t}_R$) is included in the calculations.

First we define several quantities and symbols
\begin {eqnarray} 
&&G=\frac{g^{2}m_t}{2m_W}\ctb{|{V_{tb}}|}^2 ,\\
&&A_1=\bar{u}(p_2){ \epsilon \hspace{-2mm} \slash} P_Lv(p_1) , \\
&&A_2=\bar{u}(p_2){p_3 \hspace{-3.5mm} \slash}{\hspace{1mm}} P_Lv(p_1)p_2\cdot\epsilon ,\\
&&A_3=\bar{u}(p_2){p_3 \hspace{-3.5mm} \slash} {\hspace{1mm}} P_Lv(p_1)p_1\cdot\epsilon, 
\end {eqnarray}    
where $\ctb=v_1/v_2$ is the ratio of the vaccum expectation values of the two Higgs doublets, $V_{tb}$ is the CKM matrix element, $\epsilon$ is the polarization vector of $W^+$ boson and $p_1, p_2, p_3$ are the four-vector momenta of $\bar{b}, b$ and $W^+$ respectively.

The tree-level amplitude due to Fig. 1 is 
\begin{equation}
M_0=G\frac{m_t}{p_t^2-m_t^2}A_1 ,
\end {equation}
with $p_t$$\equiv$$p_2+p_3$
and the corresponding width\cite{ma} is 
\begin {eqnarray}
{d \Gamma_0  \over d s_{\bar b } d s_b }&=&{ 1 \over 256 \pi^3 \mh^3 } \left( { 3 g^4 m_t^4 cot^2 \beta \over 4 m_W^4 \left( m_t^2 - s_{\bar b } \right)^2 } \right) \nonumber\\
&&\left[ m_W^2 \left( s_W -2 m_b^2 \right) + \left( s_{ \bar b }- m_b^2 - m_W^2 \right) \left( s_b - m_b^2 - m_W^2 \right) \right] ,
\end {eqnarray}
where $s_{\bar b } $, $s_b $ and $ s_W $ are the 4-momentum squared
transferred to the corresponding particles by $\hp$ \cite{5}.

The one-loop QCD and SUSY-QCD corrections to the process arise from the diagrams of Fig. 3 and Fig. 4, which involve virtual gluon and gluino exchanges. The self-energy corrections of external legs make no contribution to the width and we do not depict these diagrams. The amplitudes for those diagrams in Fig. 3 and Fig. 4 are
\begin{eqnarray}
&&M_a^{(g,\tilde{g})}=G{\frac{\alpha_s}{3\pi}}{\frac{1}{{(p_t^2-m_t^2)}^2}}F_1^{(g,\tilde{g})}A_1 , \\
&&M_b^{(g,\tilde{g})}=G{\frac{\alpha_s}{3\pi}}{\frac{1}{(p_t^2-m_t^2)}}[F_2^{(g, \tilde{g})}A_1+F_3^{(g,\tilde{g})}A_2] , \\
&&M_c^{(g,\tilde{g})}=G{\frac{\alpha_s}{3\pi}}{\frac{1}{(p_t^2-m_t^2)}}[F_4^{(g, \tilde{g})}A_1] , \\
&&M_d^{(g,\tilde{g})}=G{\frac{\alpha_s}{3\pi}}[F_5^{(g,\tilde{g})}A_1+F_6^{(g,\tilde{g})}A_2+F_7^{(g,\tilde{g})}A_3] , 
\end {eqnarray}
where the indices $g$ and $\tilde{g}$ refer to the gluon and gluino corrections. $F_1^{(g,\tilde{g})}-F_7^{(g,\tilde{g})}$ are functions of various kinematic invariants, which include the two-point, three-point and four-point functions as defined in ~\cite{6}. We give the explicit forms of $F_1^{(g,\tilde{g})} - F_7^{(g,\tilde{g})}$ in Appendix B. It can be seen from above that the amplitudes for the loop diagrams in Fig 4. have the same structures as their counterparts in Fig. 3.

Now we turn to the mass and wave-function renormalization due to Fig. 2(a),(b). The self-energy parts can be written in the form
\begin {eqnarray}
\label{selfg1}
&&-i\Sigma^{(g)}={{-i {\alpha_s}} \over {3 \pi}}[{{p \hspace{-2mm} \slash}} (1-2B_0(p^2,m_q^2,\lambda^2)-2B_1(p^2,m_q^2,\lambda^2))\nonumber \\
&&\mbox{\hphantom{123456}}+(4B_0(p^2,m_q^2,\lambda^2)-2)m_q] , \\
\label{selfg2}
&&-i\Sigma^{(\tilde{g})}=-i[{p \hspace{-2mm} \slash}(\Sigma_L^{(\tilde{g})}(p^2)P_L+\Sigma_R^{(\tilde{g})}(p^2)P_R)+\Sigma_s^{(\tilde{g})}(p^2)], \\
&&\Sigma_L^{(\tilde{g})}(p^2)P_L+\Sigma_R^{(\tilde{g})}(p^2)P_R={{\alpha_s}\over {3 \pi}}[2({\cos}^2\theta P_L+{\sin}^2\theta P_R)B_1(p^2,m_{\tilde{g}}^2,m_{\tilde{q_1}}^2) \nonumber\\
&&\mbox{\hphantom{12345678910111213141516}}+2({\sin}^2\theta P_L+{\cos}^2\theta P_R)B_1(p^2,m_{\tilde{g}}^2,m_{\tilde{q_2}}^2)], \\
&&\Sigma_s^{(\tilde{g})}(p^2)={{\alpha_s} \over 3 \pi}[m_{\tilde{g}}\sin2\theta(B_0(p^2,m_{\tilde g}^2,m_{\tilde q_1}^2)-B_0(p^2,m_{\tilde g^2},m_{\tilde q_2}^2))],
\end {eqnarray}
where we have introduced a gluon mass $\lambda$ to regularize the infrared divergences and $\theta$ is the mixing angle of left-right handed top-squark defined in Appendix A.

We shall use the on-shell renormalization scheme throughout the paper\cite{6.5}. From (\ref{selfg1}) and (\ref{selfg2}), we can easily derive the mass counterterms $\delta$m
\begin{eqnarray}
&&\delta m_q^{(g)}={{\alpha_s} \over {3 \pi}}m_q[2B_0(m_q^2,m_q^2,0)-2B_1(m_q^2,m_q^2,0)-1] ,\\
&&\delta m_q^{(\tilde{g})}={\frac{\alpha_s}{3 \pi}} m_q [B_1(m_q^2,m_{\tilde{g}}^2,m_{\tilde{q_1}}^2)+B_1(m_q^2,m_{\tilde{g}}^2,m_{\tilde{q_2}}^2)\nonumber \\
&&\mbox{\hphantom{1234567}}+\frac{m_{\tilde{g}}}{m_q} \sin 2 \theta(B_0(m_q^2,m_{\tilde{g}}^2,m_{\tilde{q_1}}^2)-B_0(m_q^2,m_{\tilde{g}}^2,m_{\tilde{q_2}}^2))],
\end {eqnarray}
and the wave-function renormalization constants  
\begin {eqnarray}
\delta Z_q^{(g)}&=&{\alpha_s \over {3 \pi}} [1-2B_0(m_q^2,m_q^2,\lambda^2)-2B_1(m_q^2,m_q^2,\lambda^2)\nonumber \\
&&+4 m_q^2 (DB_0(m_q^2,m_q^2,\lambda^2)-DB_1(m_q^2,m_q^2,\lambda^2))] ,\\
\delta Z_L^{(\tilde{g})}&=&{{\Sigma}}_L^{(\tilde{g})}+m_q^2 ({\dot{\Sigma}}_L^{(\tilde{g})}+{\dot{\Sigma}}_R^{(\tilde{g})})+2m_q{\dot{\Sigma}}_s^{(\tilde{g})} ,\\
\delta Z_R^{(\tilde{g})}&=&{{\Sigma}}_R^{(\tilde{g})}+m_q^2 ({\dot{\Sigma}}_L^{(\tilde{g})}+{\dot{\Sigma}}_R^{(\tilde{g})})+2m_q{\dot{\Sigma}}_s^{(\tilde{g})},
\end {eqnarray} 
with $\dot{X}=d X/d p^2$. In the above expressions we denote $B_X$(X=0,1) as the two point functions and $DB_X$ as the derivative of $B_X$ with respect to the momentum squared. We give the related two point, three point and four point functions in Appendix B. 

The counterterms for the coupling coefficients of the Higgs-top-bottom vertex and top-bottom-$W^+$boson vertex are 
\begin {eqnarray}
&&\delta_{htb}^{(g)}=-{{\delta m_t^{(g)}} \over {m_t}}+{1 \over 2} (\delta Z_t^{(g)}+\delta Z_{b}^{(g)}) ,\\
&&\delta_{tbW}^{(g)}={1 \over 2} (\delta Z_{t}^{(g)}+\delta Z_{b}^{(g)}), \\
&&\delta_{htb}^{(\tilde{g})}=-{{\delta m_t^{(\tilde {g})}} \over {m_t}}+{1 \over 2} (\delta Z_{t_R}^{(\tilde{g})}+\delta Z_{b_L}^{(\tilde{g})}) ,\\
&&\delta_{tbW}^{(\tilde{g})}={1 \over 2} (\delta Z_{t_L}^{(\tilde g)}+\delta Z_{b_L}^{(\tilde g)})  .
\end {eqnarray}
The amplitudes of the diagrams containing above counterterms will cancel out the ultraviolet divergences contained in the amplitudes for Fig. 3(b),(c) and Fig. 4(b),(c) completely.

To remove the ultraviolet divergence of the amplitude for Fig. 4(a), we introduce the following counterterm which corresponds to Fig. 1(c)
\begin {equation}
i [p \hspace{-2mm} \slash (\delta Z_{t_L}^{(\tilde{g})} P_L+\delta Z_{t_R}^{(\tilde{g})} P_R)-m_t{{\delta Z_{t_L}^{(\tilde g)}+\delta Z_{t_R}^{({\tilde g})}} \over 2}+\delta m_t^{(\tilde{g})}]  .
\end {equation}

There is another way to get the same results. If we only substract $\delta$m from the self-energy function $\Sigma$ at the mass-shell, the remaining ultraviolet divergences will disappear automatically after we add all the amplitudes for loop diagrams(including the self-energy diagrams of the external legs) and the mass counterterms related with the $\hp tb$ vertex.

The ${\cal O}(\alpha_s)$ contributions to the three-body decay width are given by the interference terms between higher order amplitudes and tree-level amplitude, {\it i.e.}
\begin{equation}
\delta \Gamma=\int_{phasespace}2 Re (\sum\limits_{\epsilon,c,\sigma} M^{(\alpha_s)}M_0^*), 
\end {equation}
where $\epsilon$ represents the W-boson polarization and c, $\sigma$ represent the color and spin of $b$ and $\bar{b}$ respectively.

The above results still contain infrared divergences which are cancelled out by the infrared divergences in real gluon emission given by Fig. 5. In our calculations of the real gluon emission contributions, we closely follow the procedure discussed in \cite{6} where an energy cutoff $\Delta$E is adopted to distinguish soft and hard gluon. The soft gluon contribution is calculated by soft gluon approximation and the hard gluon contribution is calculated by Monte Carlo methods \cite{7,8}.
The contributions of soft gluon emission of Fig. 5(a),(b) are
\begin{equation}
\label {gluon1}
{{d \Gamma^{(a)}} \over {d s_b d \bar{s_b}}}=-{{\alpha_s} \over {3 \pi^2}} I_{11}{{d \Gamma_0} \over {d s_b d s_{\bar b}}},
\end {equation}
\begin{equation}
\label {gluon2}
{{d \Gamma^{(b)}} \over {d s_b d \bar{s_b}}}=-{{\alpha_s} \over {3 \pi^2}} I_{22}{{d \Gamma_0} \over {d s_b d s_{\bar b}}},
\end {equation}
respectively, while that of the interference term between Fig. 5(a) and Fig. 5(b) is
\begin {equation}
\label {gluon3}
{{d \Gamma^{(inter)}} \over {d s_b d {\bar{s_b}}}}={2 \alpha_s \over {3 \pi^2}}I_{12}{{d \Gamma_0} \over {d s_b d {\bar{s_b}}}},
\end {equation}
where 
\begin {equation}
I_{ij}=\int\limits_{|{\bf k}|\leq \Delta E} {d^3k \over 2\omega_k} {2p_ip_j \over {(p_ik)(p_jk)}},
\end {equation}
with $\omega_k=\sqrt{{\bf k}^2+\lambda^2}$ and $\Delta E$ the cutoff parameter.
The explicit form of $I_{ij}$ is given in ~\cite{6}.
The infrared divergence terms contained in (\ref{gluon1}) (\ref{gluon2}) are the same
\begin {equation}
\label {IR1}
-{{2 \alpha_s} \over {3 \pi}} \ln{4{\Delta E}^2 \over \lambda^2}{{d \Gamma_0} \over {d s_b d {\bar{s_b}}}}.
\end {equation}
The infrared divergence terms contained in (\ref{gluon3}) is
\begin {equation}
\label {IR2}
{4 \alpha_s \over {3 \pi}}\ln {{2 p_1 \cdot p_2} \over {m_b^2}}\ln {4{\Delta E}^2 \over \lambda^2}{{d \Gamma_0} \over {d s_b d {\bar{s_b}}}}.
\end {equation}

The infrared divergences in (\ref{gluon1}) (\ref{gluon2}) will cancel out the the infrared divergences contained in $\delta Z_b^{(g)}$ and that in (\ref{gluon3}) will cancel out the infrared divergence coming from the contribution of the box diagram in Fig. 3.
This can be seen from the analytical form of $\delta Z_b^{(g)}$
\begin{equation}
\delta Z_b^{(g)}={{\alpha_s} \over {3 \pi}}[-({1 \over \epsilon}+\ln 4\pi-\gamma_E)-2\ln {\lambda^2 \over m_b^2}-\ln {\mu^2 \over m_b^2}-4],
\end {equation}
where D=4-2$\epsilon$ is the space-time dimension, $\gamma_E$ the Euler's constant and $\mu$ the 't Hooft mass parameter in the dimensional regularization scheme. 
The divergent part of the amplitude for the box diagram is contained in the folowing integral
\begin {eqnarray}     
\label {IR}
M^{IR}&=&G {\alpha_s \over {3 \pi}}{(2 \pi \mu)}^{4-D} \int {{d^D k} \over {i\pi^2}}{{\bar{u}(p_2)2m_t {p_1 \hspace{-3.5mm} \slash} {\hspace{1mm}} {\epsilon \hspace{-1.5mm} \slash}{p_2 \hspace{-3.5mm} \slash}{\hspace{0.5mm}} P_L v(p_1)} \over {(k^2-\lambda^2)({(p_2-k)}^2-m_b^2)({(p_1+k)}^2-m_b^2)(p_t^2-m_t^2)}}\nonumber \\
&=&G {\alpha_s \over {3 \pi}} {{-4m_t {p_1 \cdot p_2} \bar{u}(p_2) {\epsilon \hspace{-1.5mm} \slash} P_L v(p_1)} \over {{p_t}^2-m_t^2}}C_0(m_b^2,m_b^2,{(p_1+p_2)}^2,m_b^2,\lambda^2,m_b^2)  .
\end {eqnarray}
The analytical expression of $C_0$ can be found in ~\cite{9}.
We can thus easily obtain the infrared part contained in the contribution due to the box diagram in Fig. 3.
\begin {equation}
{d \Gamma (box) \over d s_b d s_{\bar{b}}} \sim {-4\alpha_s \over {3 \pi}} \ln {m_b^2 \over {2p_1 \cdot p_2}} \ln {\lambda^2 \over {2p_1 \cdot p_2}}{d \Gamma_0 \over d s_b d s_{\bar{b}}}.
\end {equation}
It is evident that all the infrared divergences are cancelled out completely.

At last, we consider the charged Higgs loop-corrections to the width because we are interested in the large $\cot \beta$ region where the coupling of charged Higgs to top quark and bottom quark is large. The corresponding diagrams are shown in Fig. 6. The substraction procedure is standard\cite{6.5}. We will discuss the numerical result for it in the next section.
\section {Numerical Calculations and Discussions}
\paragraph{}

We now turn to the numerical evaluation of the corrected width. We have tested the results in a number of ways. We found that the results are reliable since they don't depend on the choice of the t'Hooft mass parameters $\mu$ in the dimensional-regularization scheme and the fictitious gluon mass $\lambda$. We also choose different energy cutoffs in the calculation of soft gluon emission and find the results are independent of it to a satisfactory precision. The cancellation of $\ln m_b$ is examined too.

The whole analysis will depend on $\mh, \tb, \mu, A(=A_t), m_{\tilde{g}}, m_{\tilde{t}_L}(=m_{\tilde{b}_L})$ and $m_{\tilde{t}_R}$. For simplicity, we have assumed $m_{\tilde{t}_L}=m_{\tilde{t}_R}$ in the calculations. We take $|V_{tb}|$=1, $m_t$=180GeV, $m_b$=4.8GeV, $m_W$=80.4GeV, $\sin^2 \theta_W$=0.23,  $\alpha$=1/128, $\alpha_s=\alpha_s$(150GeV)=0.113 and $m_Z$=91.2GeV. The SUSY parameters are constained to satisfy the lighter top squark $m_{{\tilde t}_1}\geq$90GeV.

The effects of the standard QCD and SUSY-QCD corrections of (8), (9), (10), (11) can be seen in Fig. 7, in which we take ($m_{\tilde{t}_L},m_{\tilde{g}},\mu,A$)=(200,400,-300,200)GeV. The results show the QCD corrctions raise the width by about 12$\%$. The gluino corrections can raise the width by about 9$\%$ if we take the parameters listed above. 

In Fig. 8, we show the contour lines of ${{\delta \Gamma (gluino)} \over \Gamma(tree)}$ in the A-$\mu$ plane for $\tb$=1 and ($\mh,m_{\tilde{g}},m_{\tilde{b}_L}$=$m_{\tilde {t}_L}$)=(150,400,280)GeV. This correction has a strong dependence on $\mu$. It changes sign near $\mu \sim 0$GeV. We can see for the chosen values of masses it can reach about 12$\%$ when A is near $\pm$450GeV and $\mu$ is near $\pm$450GeV.

In Fig. 9, we show the dependence of ${\delta \Gamma (gluino)} \over \Gamma (tree)$ as a function of A and $m_{\tilde{t}_L}$ for $\tb$=1 and ($\mh,m_{\tilde{g}},\mu$)=(150,420,300)GeV. We can see the gluino correction can be about ~15$\%$ when A is negative and $m_{\tilde{t}_L}$ is about 100 GeV.

In Fig. 10 we show a contour plot of ${{\delta \Gamma (gluino)} \over \Gamma(tree)}$ in the $\mu$-$m_{\tilde{t}_L}$ plane for $\tb=1$ and ($m_{H^+},m_{\tilde{g}},A$)=(150,420,300)GeV. We find the corrections can reach $20\%$ when $\mu=-500 \sim -400$ and $m_{\tilde{t}_L}$ is about 100 GeV.

In Fig. 11 we show a contour plot of ${{\delta \Gamma (gluino)} \over \Gamma(tree)}$ in the $\tb$-$m_{\tilde{g}}$ plane for ($\mh$,$\mu$,$m_{\tilde{t}_L}$, A)=($150,-300,280,300$)GeV. Devided by $\tb$=1.1, the left part of the graph is similar to the right part. The correction rises when $\tb$ deviate from $\tb$=1.1. Given a $\tb$, the corrections increase up to about $m_{\tilde {g}}$=650GeV and then decrease as $m_{\tilde{g}}$ inceases.

The reason for the large contribution of $\delta \Gamma(gluino)$ is mainly because the vertex correction part of the gluino-exchange corrections is proportional to the $\hp {\tilde{t}} {\tilde{b}}$ coupling which can be enlarged greatly if the $\tilde t$-mixing parameters A and $\mu$ are large.

In Fig. 12, we show the ratio of the three-body decay width including the standard QCD corrections to the width of the two-body decay ($\hp \rightarrow \cs, \tv$), in which we can clearly see the corrected three-body decay rises sharply with increasing $\mh$. It can reach more than 8 times the size of the two-body decay in the given area. 
 
In Fig. 13, we show the ratio of $\delta \Gamma^{(\tilde g)}$, the corrections due to gluino exchanges, to the width of the two-body decay ($\hp \rightarrow  \tv$), in which we take ($m_{\tilde{t}_L},m_{\tilde{g}},\mu,A$)=($200,\\400,-300,200$)GeV. These curves indicate that the ratio tends to be smaller when $\tb$ increases while it can be large when $\tb$ is less than 1.

Finally, we have calculated the dominant terms from the Higgs loop corrections in Fig. 6. We find the Higgs corrections are relatively small, which can lower the width by about 0.6$\% \cot^2 \beta$ when $\mh$=140GeV, and the ratio of the Higgs corrections to the tree-level contribution decreses when $\mh$ increases. Therefore, the Higgs corrections can be neglected if $\tb$ is not too small.

In Fig. 14, we show the ratio of $\Gamma_0+\delta \Gamma^{(g)}+\delta \Gamma^{(\hp)}$(the width including the standard QCD and Charged Higgs correcions) to the width of the two-body decay ($\hp \rightarrow  \tv$). If we measure the three channels $\hp \rightarrow \tau^+\nu, \hp \rightarrow c \bar{s}$ and $\hp \rightarrow W^+ b \bar{b}$ simultaneously and obtain $\tb$ of the two-Higgs doublet model of type II from $\Gamma_{c\bar{s}}/\Gamma_{\tau^+\nu}$, we can check the validity of the two-Higgs doublet model of type II without supersymmetry by comparing the experimental data for $\Gamma_3/\Gamma_{\tau^+ \nu}$ with theoretical results like those in Fig. 14.

In summary, we have performed complete calculations of the ${\cal O}(\alpha_s)$ standard QCD and SUSY-QCD corrections to the width of $\hp \rightarrow b{\bar {b}} W^+$. We have found that the QCD corrections raise the width by about 12$\%$ and the SUSY-QCD corrections can be comparable to or even larger than the standard QCD corrections and change signs as $\mu$ varies. This provides an effective way to distinguish the two-Higgs doublet model of type II from the MSSM.

\newpage
\begin{center}
{\bf \Huge  Appendix A }
\end{center}
\newcounter{num}
\setcounter{num}{1}
\setcounter{equation}{0}
\def\theequation{\Alph{num}.\arabic{equation}}
%\paragraph{}
%\mbox{\hphantom{1}}
In this appendix we list some relevant pieces of the SUSY Lagrangian.\\
The charged Higgs boson coupling to top and bottom quarks is given by
\begin{equation}
{\cal L}={g V_{tb}\over \sqrt{2} m_W}H^+[m_t \ctb \bar{t}P_L b+m_b \tb \bar{b}P_R t]+h.c.,
\end {equation}
where $P_{L,R}={1 \over 2}(1\mp \gamma_5)$ are the chiral projector operators, $\ctb=v_1/v_2$ is the ratio of the vaccum expectation values of the two Higgs doublets and $V_{tb}$ is the CKM matrix element.\\ 
The squark couplings to the charged Higgs, gluino and $W^+$ boson are given by
\begin{eqnarray}
&&{\cal L}_{{\tilde t} \tilde{b} {H}}={-g \over \sqrt{2} m_W}[(m_W^2 \sin2 \beta-m_t^2 \cot \beta-m_b^2 \tan \beta)Z^b_{1j} Z^t_{1i}-{2m_tm_b \over \sin2 \beta}Z^b_{2j}Z^t_{2i},\nonumber \\
&&\mbox{\hphantom{123}}+m_t(A_t^* \cot \beta-\mu)Z^t_{2i}Z^b_{1j}+m_b(A_b \tan \beta-\mu^*)Z^b_{2j}Z^t_{1i}]V_{tb}^*{\tilde t}_i{\tilde b}_j^*H^-+h.c.,\nonumber
\\ 
&&\\
&&{\cal L}_{{\tilde t} t {\tilde g}}=g_s \sqrt{2} {{\tilde t}_i^*} {T^a} {\bar{{\tilde g}}^a}[-{Z^{t*}_{1i}}P_L+{Z^{t*}_{2i}}P_R]t+h.c., \\
&&{\cal L}_{{\tilde b}b{\tilde{g}}}=g_s \sqrt{2} {{\tilde b}_i^*} {T^a} {\bar{{\tilde g}}^a}[-Z^{b*}_{1i}P_L+Z^{b*}_{2i}P_R]b+h.c.,\\
&&{\cal L}_{{\tilde b}{\tilde t}W}={-g \over \sqrt{2}}Z_{1i}^b Z_{1j}^t V_{tb}({\tilde {b}_i^*}{\stackrel{\leftrightarrow}{\partial}}^\mu \tilde{t}_j) W_{\mu}^-+h.c,
\end {eqnarray}
respectively, in which $g_s$ is the QCD coupling constant, g is the EW coupling constant and $T^a$ is the matrices of the SU(3) generators in the 3 representation.
\setcounter{equation}{5}
\def\theequation{\Alph{num}.\arabic{equation}}
\begin {equation}
Z^{t}=
\left(\begin{array}{cc}
\cos\theta & -\sin\theta\\
\sin\theta & \cos\theta
\end {array}
\right)
\end {equation}
and $Z^b$ are orthogonal matrices which diagonalize the mass matrices $M_{\tilde{t}}^2$ and $M_{\tilde{b}}^2$ of the squarks.
%The mass matrices of the squarks are
\begin {equation}
M_{\tilde{t}}^2=\left(\begin{array}{cc} m_{\tilde{t}_L}^2+m_{t}^2+0.35(m_Z^2\cos2\beta) & -m_t(A_t+\mu\cot\beta)\\ -m_t(A_t+\mu\cot\beta) & m_{{\tilde t}_R}^2+m_{t}^2+0.15(m_Z^2\cos2\beta)\end {array} \right),\\
\end {equation}
\begin {equation}
M_{\tilde{b}}^2=\left(\begin{array}{cc}
m_{\tilde{b}_L}^2+m_{b}^2-0.42(m_Z^2\cos2\beta) & -m_b(A_b+\mu\tb)\\
-m_b(A_b+\mu\tb) & m_{{\tilde b}_R}^2+m_{b}^2-0.08(m_Z^2\cos2\beta)
\end {array}\right),
\end {equation}
when we neglect the mass of bottom-quark, the last mass matrix reduces to
\begin {equation}
M_{\tilde{b}}^2=\left(\begin{array}{cc}
m_{\tilde{b}_L}^2-0.42(m_Z^2\cos2\beta) & \\
& m_{\tilde{b}_R}^2-0.08(m_Z^2\cos2\beta)
\end {array}\right).
\end {equation}

\newpage
\textwidth 152.5mm
\textheight 210mm
\topmargin -15mm
\oddsidemargin 2mm
\evensidemargin 2mm
\begin{center}
{\bf \Huge  Appendix B }
\end{center}
\setcounter{num}{2}
\setcounter{equation}{0}
\def\theequation{\Alph{num}.\arabic{equation}}
In below expressions, $m_{{\tilde b}}=m_{{\tilde b}_1}$ and $Z_{ij}(i,j=1,2)$ represent the elements of matrix $Z^{{t}}$ defined in Appendix A.

\begin {eqnarray}
&&F_1^{(g)}=-2m_t^3+2m_t^3B_0(0,m_t^2,0)+2m_t(m_t^2+s_{\bar{b}})B_0(s_{\bar{b}},m_t^2,0) ,\\
&&F_2^{(g)}=-m_t[ 1-2B_0(m_W^2,0,m_t^2)+2(m_W^2-s_{\bar{b}})C_0(m_W^2,s_{\bar{b}},m_b^2,m_b^2,m_t^2,0)\nonumber\\
&&\mbox{\hphantom{123456}}+2(m_W^2-2s_{\bar{b}})C_1(s_{\bar{b}},m_W^2,m_b^2,0,m_
t^2,m_b^2)\nonumber \\
&&\mbox{\hphantom{123456}}+2(m_W^2-s_{\bar{b}})C_2(s_{\bar{b}},m_W^2,m_b^2,0,m_t^2,m_b^2)\nonumber \\
&&\mbox{\hphantom{123456}}+4C_{00}(s_{\bar{b}},m_W^2,m_b^2,0,m_t^2,m_b^2)],\\
&&F_3^{(g)}=4m_t[C_1(s_{\bar{b}},m_W^2,m_b^2,0,m_t^2,m_b^2)+C_{11}(s_{\bar{b}},m_W^2,m_b^2,0,m_t^2,m_b^2)+\nonumber \\
&&\mbox{\hphantom{123456}}C_{12}(s_{\bar{b}},m_W^2,m_b^2,0,m_t^2,m_b^2)],\\
&&F_4^{(g)}=-2m_t[1-2B_0(\mh^2,0,m_t^2)\nonumber \\
&&\mbox{\hphantom{123456}}-(m_W^2-s_b-s_W)C_0(s_{\bar{b}},\mh^2,m_b^2,0,m_t^2,m_b^2)\nonumber \\
&&\mbox{\hphantom{123456}}-(m_W^2+s_{\bar{b}}-s_b-s_W)C_1(s_{\bar{b}},\mh^2,m_b^2,0,m_t^2,m_b^2)\nonumber \\
&&\mbox{\hphantom{123456}}-(m_W^2-s_b-s_W)C_2(s_{\bar{b}},\mh^2,m_b^2,0,m_t^2,m_b^2)],\\
&&F_5^{(g)}=2m_tC_0(m_W^2,s_W,\mh^2,m_t^2,m_b^2,m_b^2)\nonumber\\
&&\mbox{\hphantom{123456}}-2m_ts_WD_0(m_b^2,m_W^2,\mh^2,m_b^2,s_{\bar{b}},s_W,0,m_b^2,m_t^2,m_b^2)\nonumber \\
&&\mbox{\hphantom{123456}}-2m_t s_WD_1(m_b^2,m_W^2,\mh^2,m_b^2,s_{\bar{b}},s_W,0,m_b^2,m_t^2,m_b^2)\nonumber \\
&&\mbox{\hphantom{123456}}+2m_t(s_{\bar{b}}-s_b-s_W)D_2(m_b^2,m_W^2,\mh^2,m_b^2,s_{\bar{b}},s_W,0,m_b^2,m_t^2,m_b^2)\nonumber \\
&&\mbox{\hphantom{123456}}-2m_t s_WD_3(m_b^2,m_W^2,\mh^2,m_b^2,s_{\bar{b}},s_W,0,m_b^2,m_t^2,m_b^2)\nonumber \\
&&\mbox{\hphantom{123456}}-4m_tD_{00}(m_b^2,m_W^2,\mh^2,m_b^2,s_{\bar{b}},s_W,0,m_b^2,m_t^2,m_b^2),\\
&&F_6^{(g)}=-4m_tD_2(m_b^2,m_W^2,\mh^2,m_b^2,s_{\bar{b}},s_W,0,m_b^2,m_t^2,m_b^2)\nonumber \\
&&\mbox{\hphantom{123456}}-4m_tD_{12}(m_b^2,m_W^2,\mh^2,m_b^2,s_{\bar{b}},s_W,0,m_b^2,m_t^2,m_b^2)\nonumber \\
&&\mbox{\hphantom{123456}}-4m_tD_{22}(m_b^2,m_W^2,\mh^2,m_b^2,s_{\bar{b}},s_W,0,m_b^2,m_t^2,m_b^2), \\
&&F_7^{(g)}=4m_tD_2(m_b^2,m_W^2,\mh^2,m_b^2,s_{\bar{b}},s_W,0,m_b^2,m_t^2,m_b^2)\nonumber \\
&&\mbox{\hphantom{123456}}+4m_tD_{23}(m_b^2,m_W^2,\mh^2,m_b^2,s_{\bar{b}},s_W,0,m_b^2,m_t^2,m_b^2),\\
&&F_1^{(\tilde{g})}=m_tB_0(0,m_{\tilde{g}}^2,m_{\tilde{t}_i}^2)(m_{\tilde{g}}^2-m_{\tilde{t}_i}^2)(Z_{1i}^2+Z_{2i}^2)-B_0(s_{\bar{b}},m_{\tilde{g}}^2,m_{\tilde{t}_i}^2)(m_{\tilde{g}}^2m_tZ_{1i}^2\nonumber \\
&&\mbox{\hphantom{123456}}+m_ts_{\bar{b}}Z_{1i}^2-m_tm_{\tilde{t}_i}^2Z_{1i}^2-2m_{\tilde{g}}m_t^2Z_{1i}Z_{2i}-2m_{\tilde{g}}s_{\bar{b}}Z_{1i}Z_{2i}\nonumber \\
&&\mbox{\hphantom{123456}}+m_{\tilde{g}}^2m_tZ_{2i}^2+m_ts_{\bar{b}}Z_{2i}^2-m_tm_{\tilde{t}_i}^2Z_{2i}^2),\\
&&F_2^{(\tilde{g})}=4m_t[C_{00}(m_b^2,m_W^2,s_{\bar{b}},m_{\tilde{g}}^2,m_{\tilde{b}}^2,m_{\tilde{t}_i}^2)Z_{1i}^2],\\
&&F_3^{(\tilde{g})}=4Z_{1i}[m_tC_{12}(m_b^2,m_W^2,s_{\bar{b}},m_{\tilde{g}}^2,m_{\tilde{b}}^2,m_{\tilde{t}_i}^2)Z_{1i}\nonumber \\
&&\mbox{\hphantom{123456}}+m_tC_{22}(m_b^2,m_W^2,s_{\bar{b}},m_{\tilde{g}}^2,m_{\tilde{b}}^2,m_{\tilde{t}_i}^2)\nonumber \\
&&\mbox{\hphantom{123456}}+m_{\tilde{g}}C_0(m_W^2,s_{\bar{b}},m_b^2,m_{\tilde{b}}^2,m_{\tilde{t}_i}^2,m_{\tilde{g}}^2)Z_{2i}\nonumber \\
&&\mbox{\hphantom{123456}}+m_{\tilde{g}}C_1(m_b^2,m_W^2,s_{\tilde{b}},m_{\tilde{g}}^2,m_{\tilde{b}}^2,m_{\tilde {t}_i}^2)Z_{2i}\nonumber \\
&&\mbox{\hphantom{123456}}+C_2(m_b^2,m_W^2,s_{\tilde{b}},m_{\tilde{g}}^2,m_{\tilde{b}}^2,m_{\tilde {t}_i}^2)(m_{\tilde{g}}Z_{2i}+m_tZ_{1i})],\\
&&F_4^{(\tilde{g})}={-2 \over m_t \ctb}[s_{\bar{b}}C_1(s_{\bar{b}},\mh^2,m_b^2,m_{\tilde{g}}^2,m_{\tilde{t}_i}^2,m_{\tilde{b}}^2)Z_{1i}\nonumber\\
&&\mbox{\hphantom{123456}}+m_{\tilde{g}}m_tC_0(s_{\bar{b}},\mh^2,m_b^2,m_{\tilde{g}}^2,m_{\tilde{t}_i}^2,m_{\tilde{b}}^2)Z_{2i}]\nonumber \\
&&\mbox{\hphantom{123456}}[(m_W^2 \sin 2\beta-m_t^2\ctb)Z_{1i}+m_t(A\ctb-\mu)Z_{2i}],\\
&&F_5^{(\tilde{g})}={-4 \over m_t\cot\beta }D_{00}(m_b^2,m_W^2,\mh^2,m_b^2,s_{\bar{b}},s_W,m_{\tilde{g}}^2,m_{\tilde{b}}^2,m_{\tilde{t}_i}^2,m_{\tilde{b}}^2)\nonumber \\
&&\mbox{\hphantom{123456}}Z_{1i}[(m_W^2\sin 2 \beta-m_t^2\ctb)Z_{1i}+m_t(A\ctb-\mu)Z_{2i}],\\
&&F_6^{(\tilde{g})}={-4 \over m_t \ctb}[D_2(m_b^2,m_W^2,\mh^2,m_b^2,s_{\bar{b}},s_W,m_{\tilde{g}}^2,m_{\tilde{b}}^2,m_{\tilde{t}_i}^2,m_{\tilde{b}}^2)\nonumber \\
&&\mbox{\hphantom{123456}}Z_{1i}((m_W^2\sin 2 \beta-m_t^2\ctb)Z_{1i}+m_t(A\ctb-\mu)Z_{2i})\nonumber \\
&&\mbox{\hphantom{123456}}+D_{12}(m_b^2,m_W^2,\mh^2,m_b^2,s_{\bar{b}},s_W,m_{\tilde{g}}^2,m_{\tilde{b}}^2,m_{\tilde{t}_i}^2,m_{\tilde{b}}^2)Z_{1i}\nonumber \\
&&\mbox{\hphantom{123456}}((m_W^2\sin 2 \beta-m_t^2\ctb )Z_{1i}+m_t(A\ctb-\mu)Z_{2i})\nonumber \\
&&\mbox{\hphantom{123456}}+D_{22}(m_b^2,m_W^2,\mh^2,m_b^2,s_{\bar{b}},s_W,m_{\tilde{g}}^2,m_{\tilde{b}}^2,m_{\tilde{t}_i}^2,m_{\tilde{b}}^2)Z_{1i}\nonumber \\
&&\mbox{\hphantom{123456}}((m_W^2\sin 2 \beta-m_t^2\ctb)Z_{1i}+m_t(A\ctb-\mu)Z_{2i})],\\
&&F_7^{(\tilde{g})}={4 \over m_t \ctb}D_{23}(m_b^2,m_W^2,\mh^2,m_b^2,s_{\bar{b}},s_W,m_{\tilde{g}}^2,m_{\tilde{b}}^2,m_{\tilde{t}_i}^2,m_{\tilde{b}}^2)\nonumber\\
&&\mbox{\hphantom{123456}}Z_{1i}[(m_W^2\sin 2 \beta-m_t^2\ctb)Z_{1i}+m_t(A\ctb-\mu)Z_{2i}].
\end {eqnarray}
\begin {flushleft}
The relevant scalar fuctions are defined as follows
\setcounter{num}{2}
\setcounter{equation}{14}
\def\theequation{\Alph{num}.\arabic{equation}}
\begin {eqnarray}
&&B_0(p_1^2,m_0^2,m_1^2)={(i \pi^2) }^{-1}{(2 \pi \mu)}^{4-D} \int d^Dq {[(q^2-m_0^2)((q+p_1)^2-m_1^2)]}^{-1},\\
&&C_0(p_1^2,p_{12},p_2^2,m_0^2,m_1^2,m_2^2)\nonumber \\
&&\mbox{\hphantom{123}}={(i \pi^2) }^{-1}{(2 \pi \mu)}^{4-D} \int d^Dq {[(q^2-m_0^2)((q+p_1)^2-m_1^2)((q+p_2)^2-m_2^2)]}^{-1},\nonumber \\
\\
&&D_0(p_1^2,p_{12},p_{23},p_3^2,p_2^2,p_{13},m_0^2,m_1^2,m_2^2,m_3^2)\nonumber \\
&&\mbox{\hphantom{123}}={{(i \pi^2) }^{-1}{(2 \pi \mu)}^{4-D}}\nonumber \\
&&\mbox{\hphantom{123}}\int d^Dq {[(q^2-m_0^2)((q+p_1)^2-m_1^2)((q+p_2)^2-m_2^2)((q+p_3)^2-m_3^2)]}^{-1}
,\\
&&DB_0(p_1^2,m_0^2,m_1^2)={\partial B_0(p_1^2,m_0^2,m_1^2) \over \partial {p_1^2}},
\end {eqnarray}
in which $p_{ij}={(p_i-p_j)}^2$.

The definitions of the tensor-integrals and the relevant decomposions are given below
\setcounter{num}{2}
\setcounter{equation}{19}
\def\theequation{\Alph{num}.\arabic{equation}}
\begin{eqnarray}
T_{\mu_1 \cdot \cdot \cdot \mu_p}(p_1,\cdot \cdot \cdot,p_{N-1},m_0,\cdot \cdot \cdot ,m_{N-1})={{(2 \pi \mu)}^{4-D} \over {i \pi^2}} \int d^Dq {q_{\mu_1} \cdot \cdot \cdot q_{\mu_n} \over D_0 D_1 \cdot \cdot \cdot D_{N-1}}, 
\end {eqnarray}
with the denominator factors $D_0=q^2-m_0^2,D_i={(q+p_i)}^2-m_i^2$  (i=1,$\cdot \cdot \cdot$,N-1)
\setcounter{num}{2}
\setcounter{equation}{20}
\def\theequation{\Alph{num}.\arabic{equation}}
\begin{eqnarray}
&&B_{\mu}={p_1}_{\mu}B_1,\\
&&C_{\mu}={p_1}_{\mu}C_1+{p_2}_{\mu}C_2=\sum\limits_{i=1}^2 p_{i\mu}C_i,\\
&&C_{\mu \nu}=g_{\mu\nu}C_{00}+{p_1}_{\mu}{p_1}_{\nu}C_{11}+{p_2}_{\mu}{p_2}_{\nu}C_{22}+({p_1}_{\mu}{p_2}_{\nu}+{p_2}_{\mu}{p_1}_{\nu})C_{12},\\
&&\mbox{\hphantom{123}}=g_{\mu\nu}C_{00}+\sum\limits_{i,j=1}^2{p_i}_{\mu}{p_j}_{\nu}C_{ij},\\
&&D_{\mu}=\sum\limits_{i=1}^3 p_{i\mu}D_i,\\
&&D_{\mu\nu}=g_{\mu\nu}D_{00}+\sum\limits_{i,j=1}^3 {p_i}_{\mu}{p_j}_{\nu}D_{ij}.
\end {eqnarray}

\end {flushleft}

\newpage
\topmargin -15mm
\oddsidemargin 6.5mm
\evensidemargin 6.5mm
\newcounter{fig}
\section *{Figure Captions}
\begin{list}{{\bf FIG. \arabic{fig}}}{\usecounter{fig}}
\item
Tree-level diagram for the three-body decay of the charged Higgs ${\hp \rightarrow W^+ b \bar{b}}$.
\item
Diagrams relevant for  wave-function and mass renormalization in the calculations of the ${\cal O}(\alpha_s)$ QCD and SUSY-QCD corrections to the width of the three-body decay ${\hp \rightarrow W^+ b \bar{b}}$.
\item
Diagrams relevant for the calculation of the ${\cal O}(\alpha_s)$ QCD corrections to the width of the three-body decay ${\hp \rightarrow W^+ b \bar{b}}$.
\item
Diagrams relevant for the calculation of the ${\cal O}(\alpha_s)$ gluino corrections to the width of the three-body decay ${\hp \rightarrow W^+ b \bar{b}}$.
\item
Diagrams for real-gluon emission ${\hp \rightarrow W^+ b \bar{b}}g$.
\item 
Diagrams relevant for the charged Higgs corrections to the width of the three-body decay ${\hp \rightarrow W^+ b \bar{b}}$.
\item
Comparision of the tree-level width with those including the standard QCD or gluino corrections . The parameters taken in gluino corrections are ($m_{\tilde{t}_L},m_{\tilde g},\mu,\\A$)=($200, 400,-300,200$). 
\item
Contour lines of $\delta \Gamma(gluino)/\Gamma(tree)$ for ($\mh,m_{\tilde{g}},m_{t_L}$)=($150,400,280$)GeV and $\tb$=1 in the A-$\mu$ plane.
\item
Contour lines of $\delta \Gamma(gluino)/\Gamma(tree)$ for ($\mh,m_{\tilde{g}},\mu$)=($150,420,300$)GeV and $\tb$=1 in the $A$-$m_{\tilde{t}_L}$ plane.
\item
Contour lines of $\delta \Gamma(gluino)/\Gamma(tree)$ for ($m_{H^+},m_{\tilde{g}}$,A)=($150,420,300$)GeV and $\tb$=1 in the $\mu$-$m_{\tilde{t}_L}$ plane.
\item
Contour lines of $\delta \Gamma(gluino)/\Gamma(tree)$ for ($\mh,\mu,m_{\tilde{t}_L},A$)=($150,-300,280,\\300$)GeV in the $\tb$-$m_{\tilde {g}}$ plane.
\item
The ratio of the three-body decay width including the standard QCD corrections to the width of the two-body decay ($\hp \rightarrow \cs, \tv$) for $\mh$=150GeV.
\item
The ratio of the gluino corrections to the three-body decay width to the width of the two-body decay ($\hp \rightarrow  \tv$) for ($\mh,m_{\tilde{t}_L},m_{\tilde{g}},\mu,A$)=($150,200,400,\\-300,200$)GeV.
\item
The ratio of the the three-body decay width including the standard QCD and charged Higgs corrections to the width of the two-body decay ($\hp \rightarrow  \tv$) for $\mh$=150GeV. 
\end{list}
\end {document}
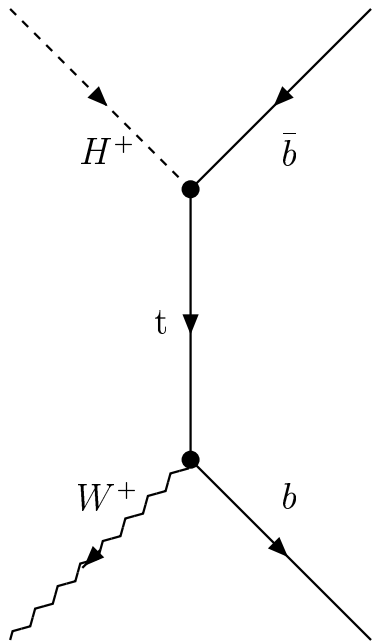
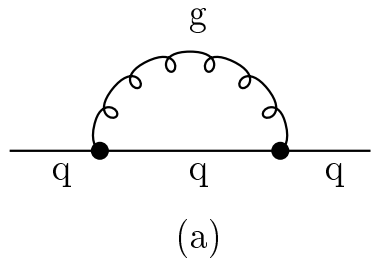
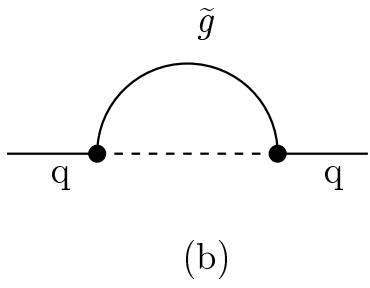
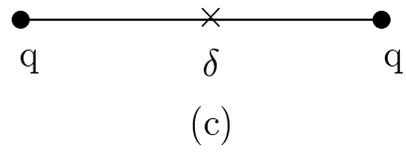
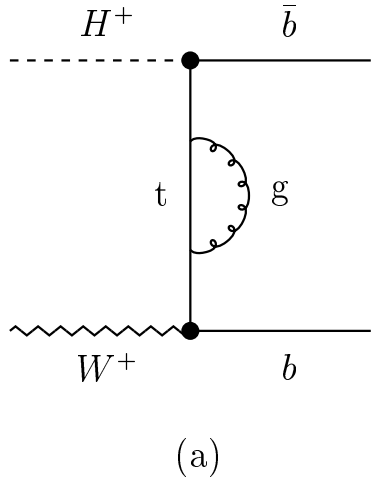
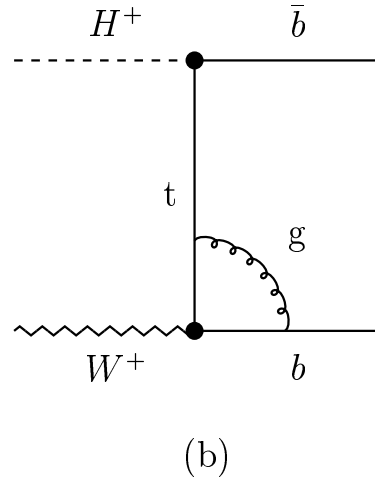
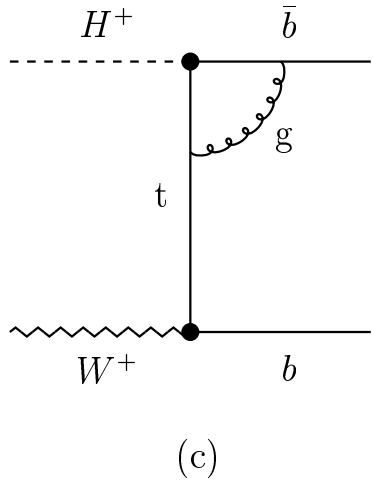
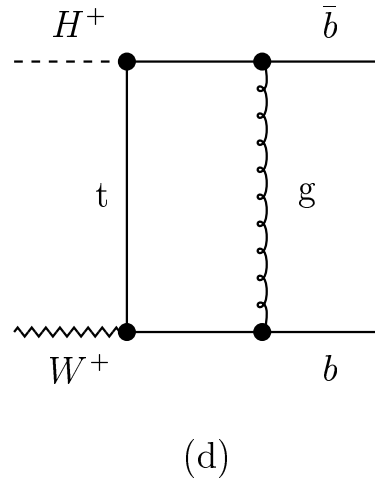
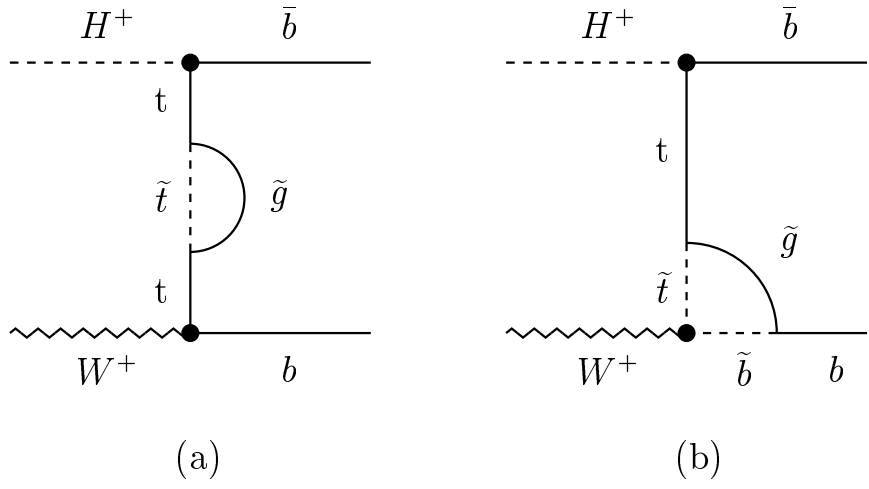
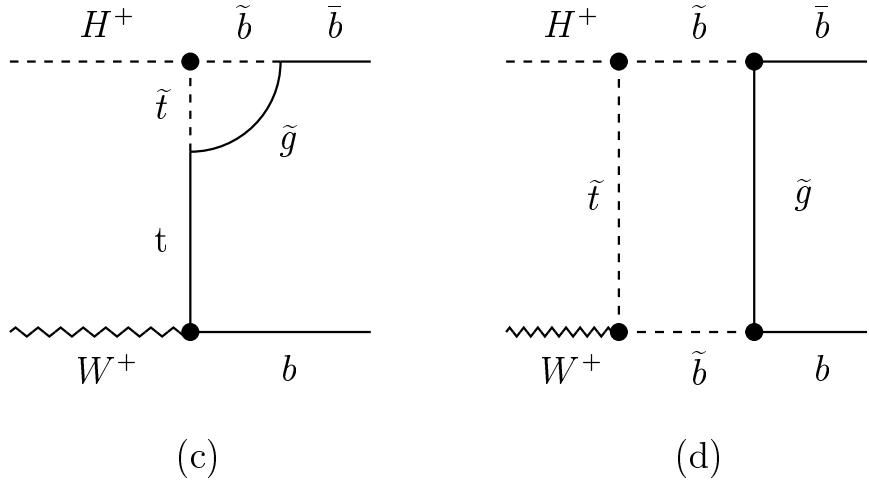
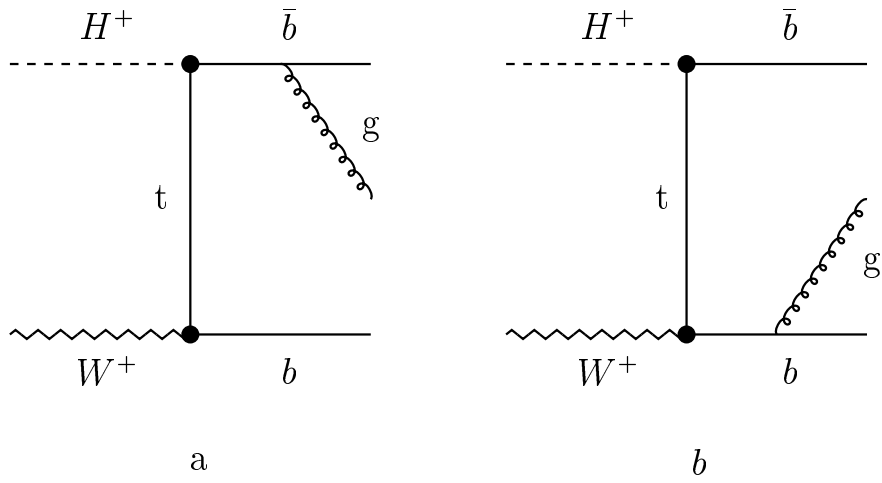
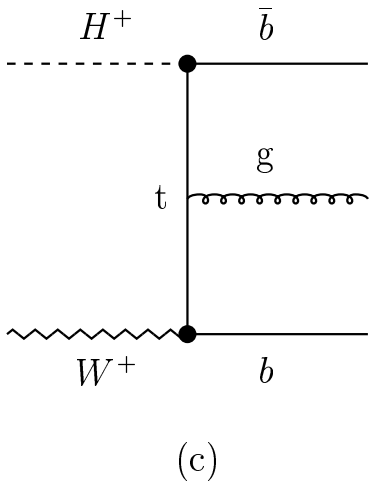
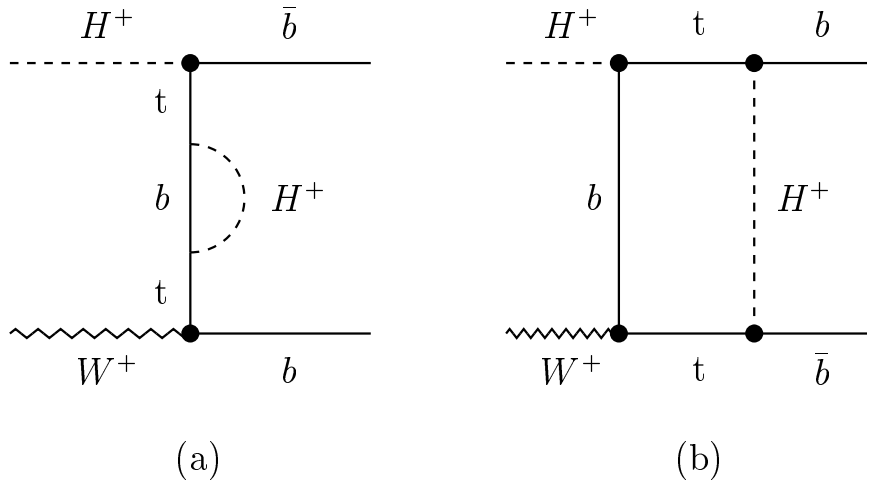
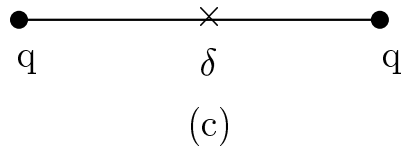
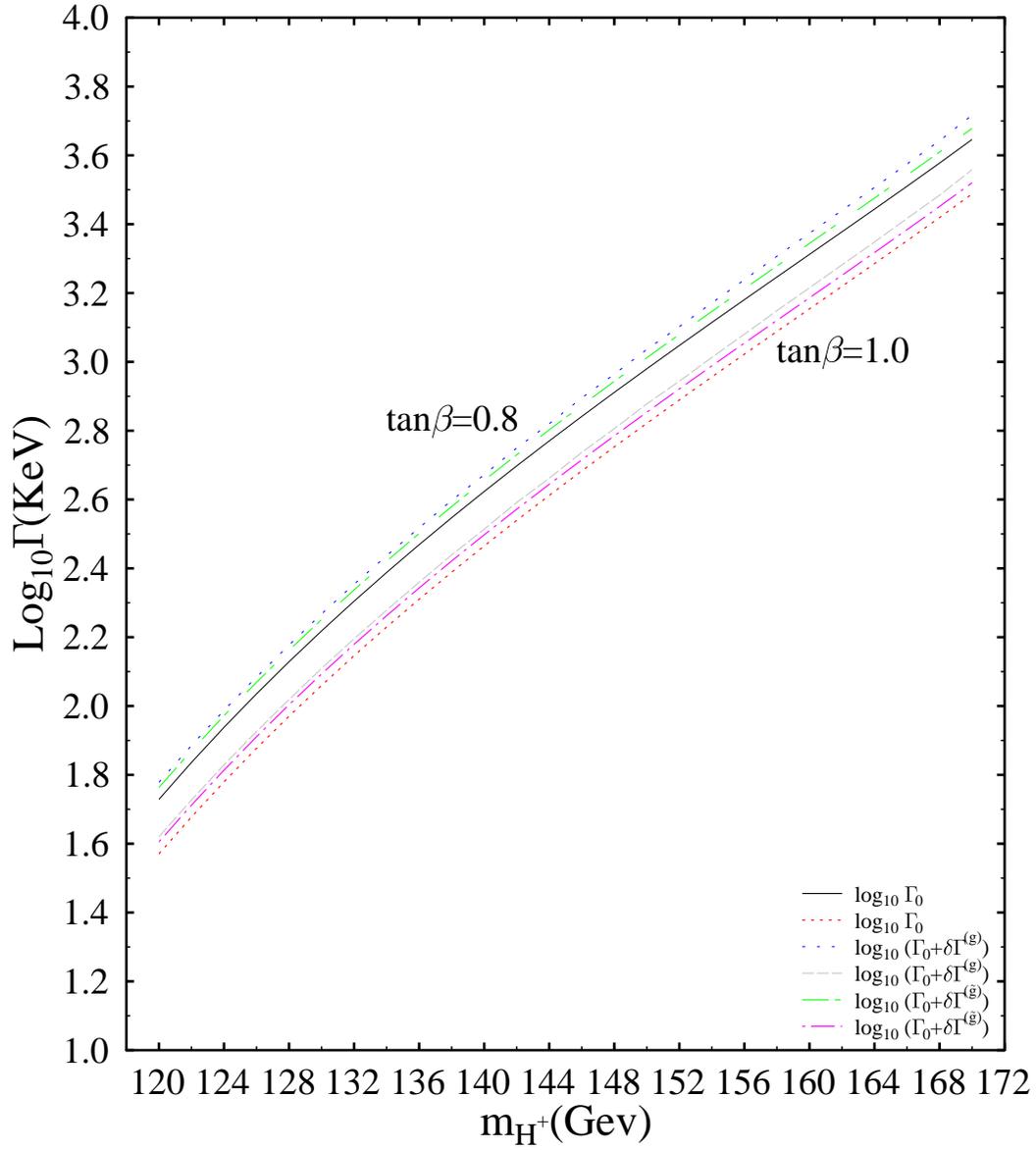
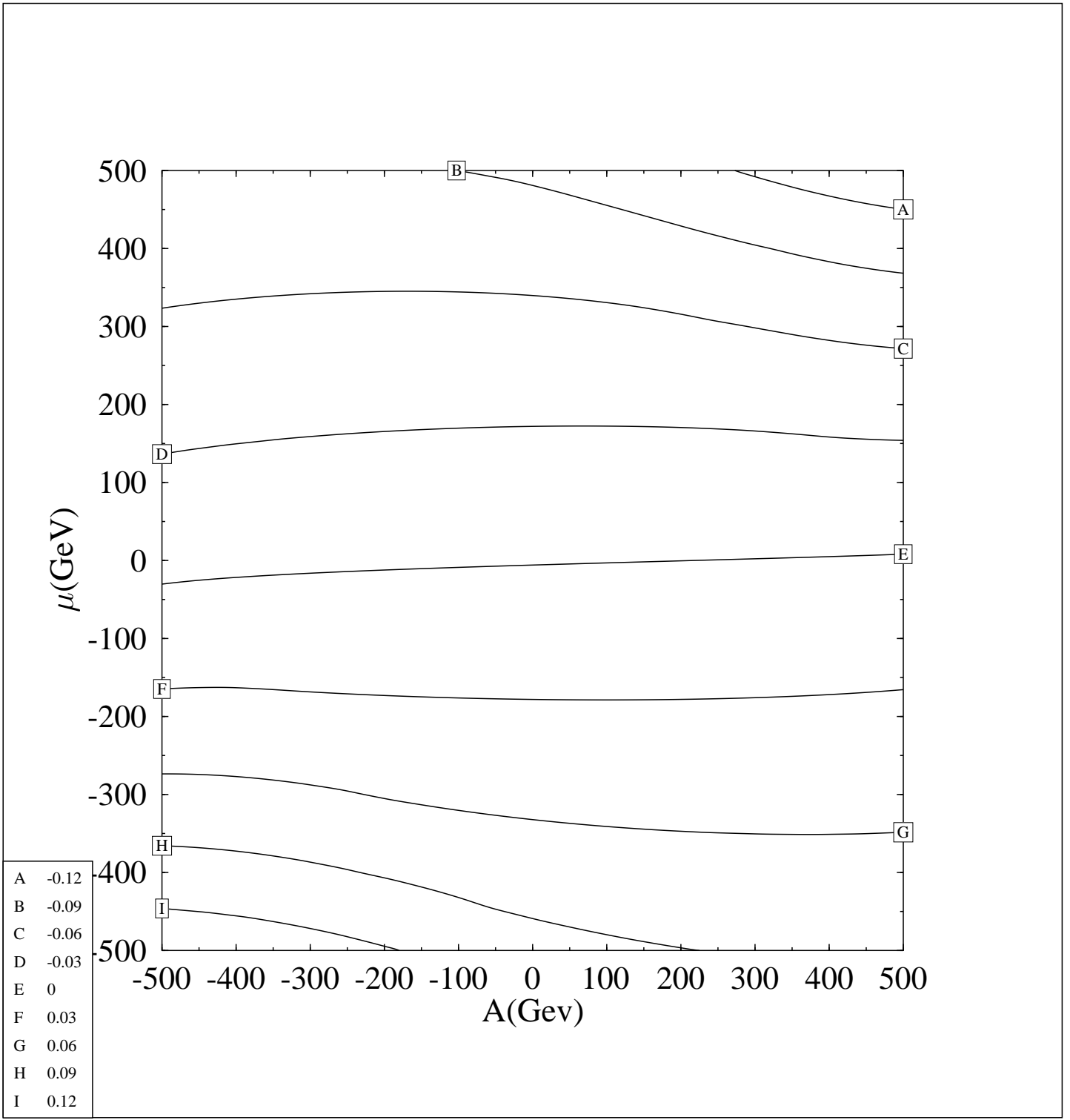
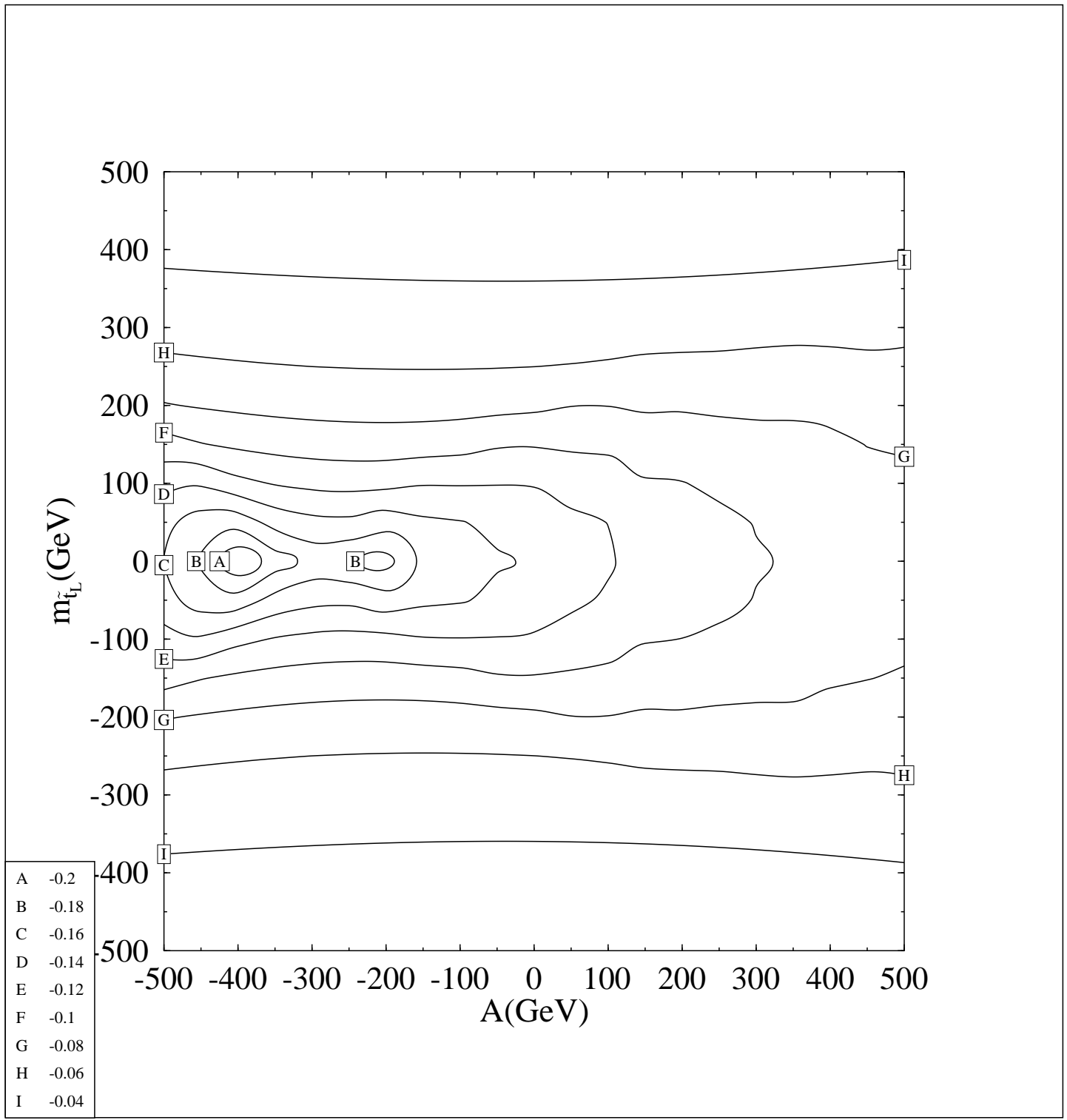
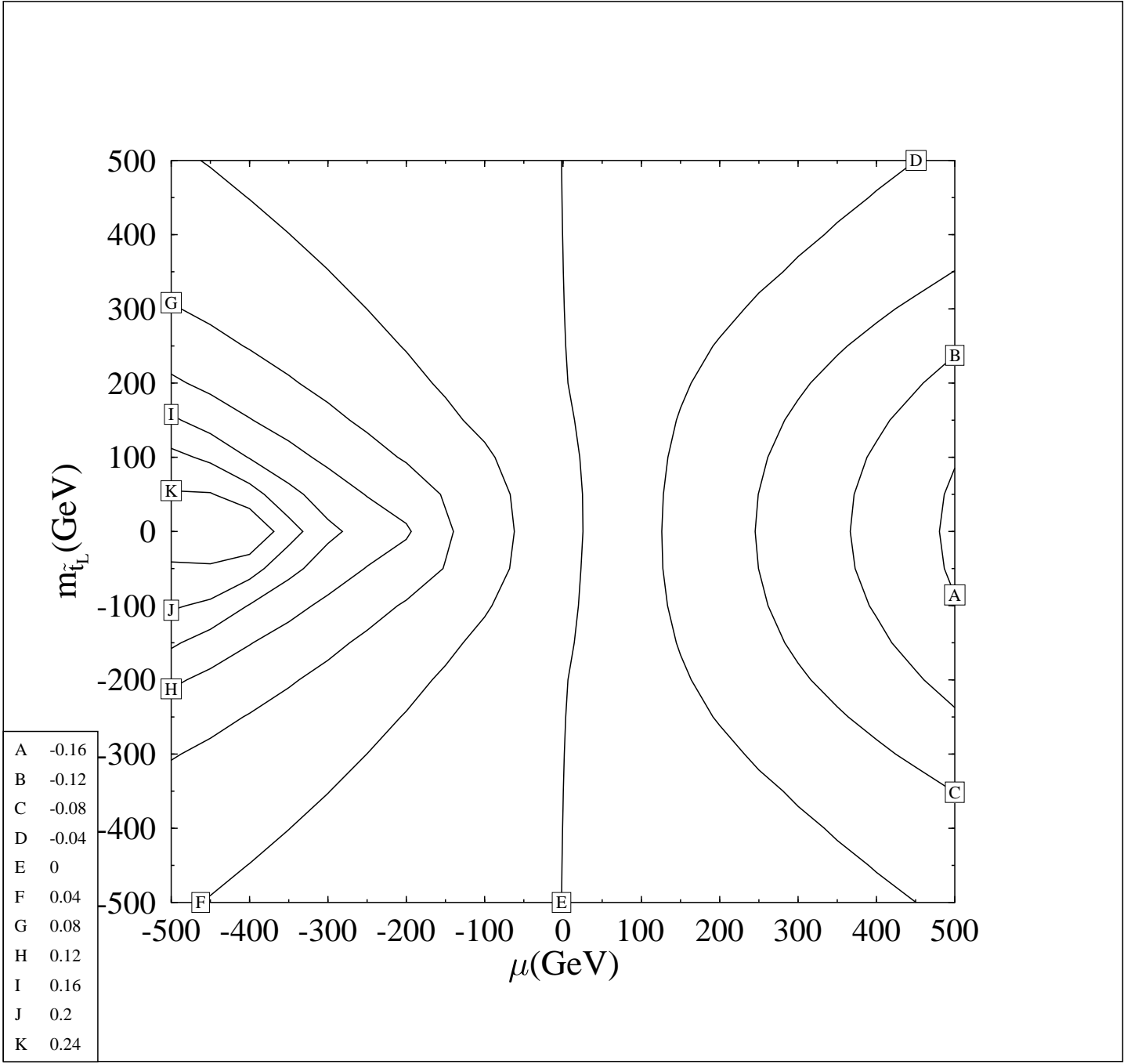
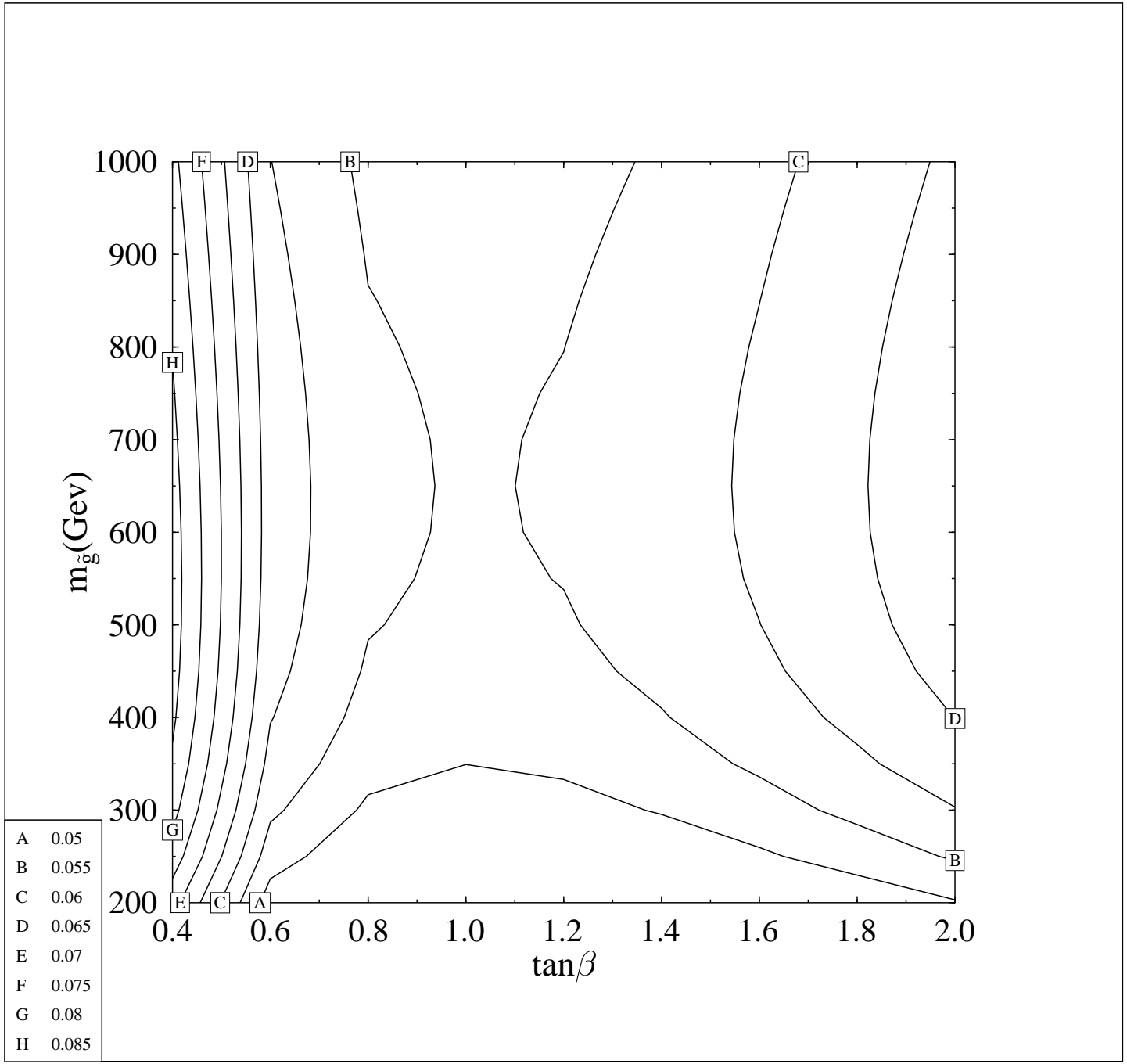
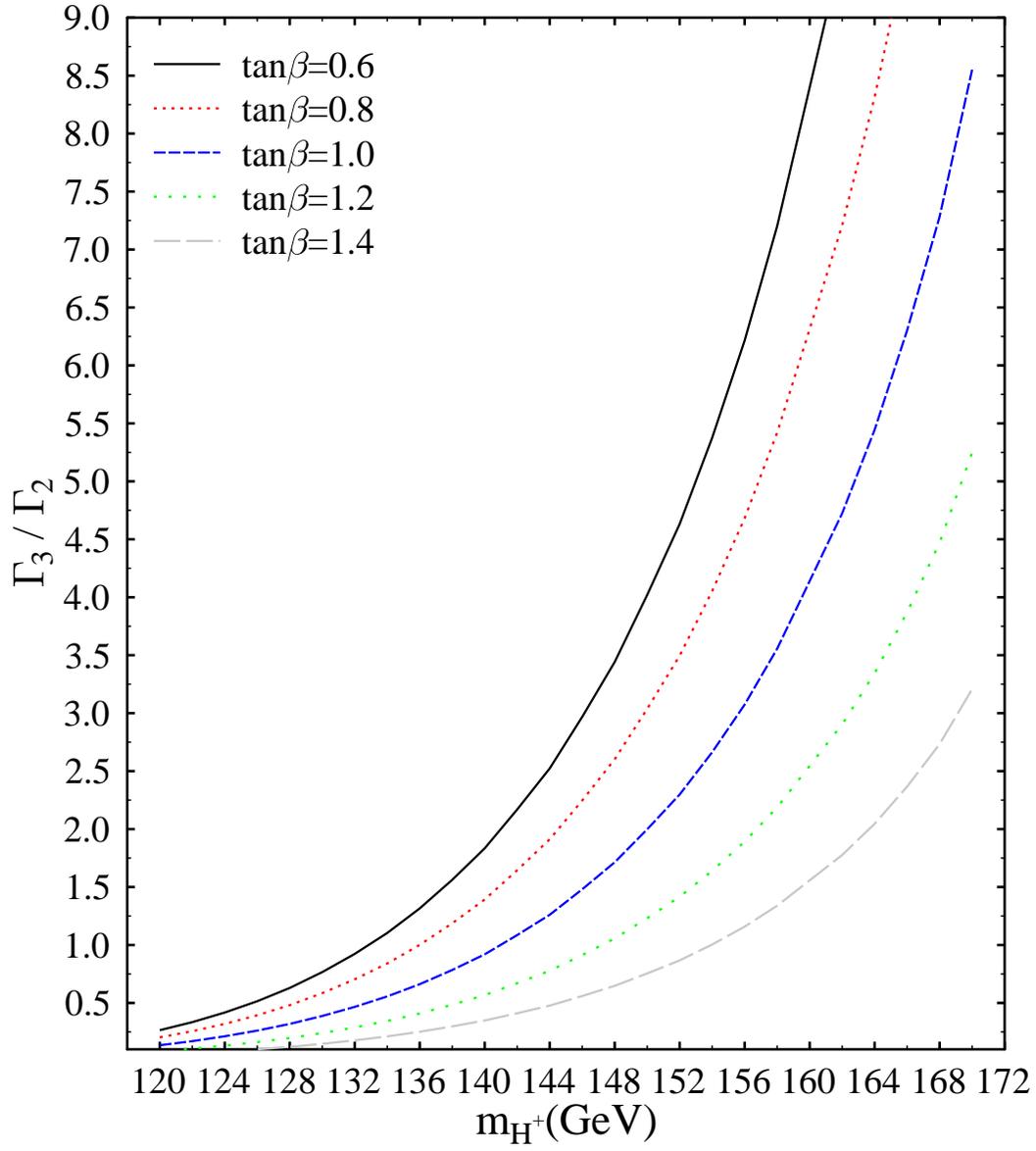
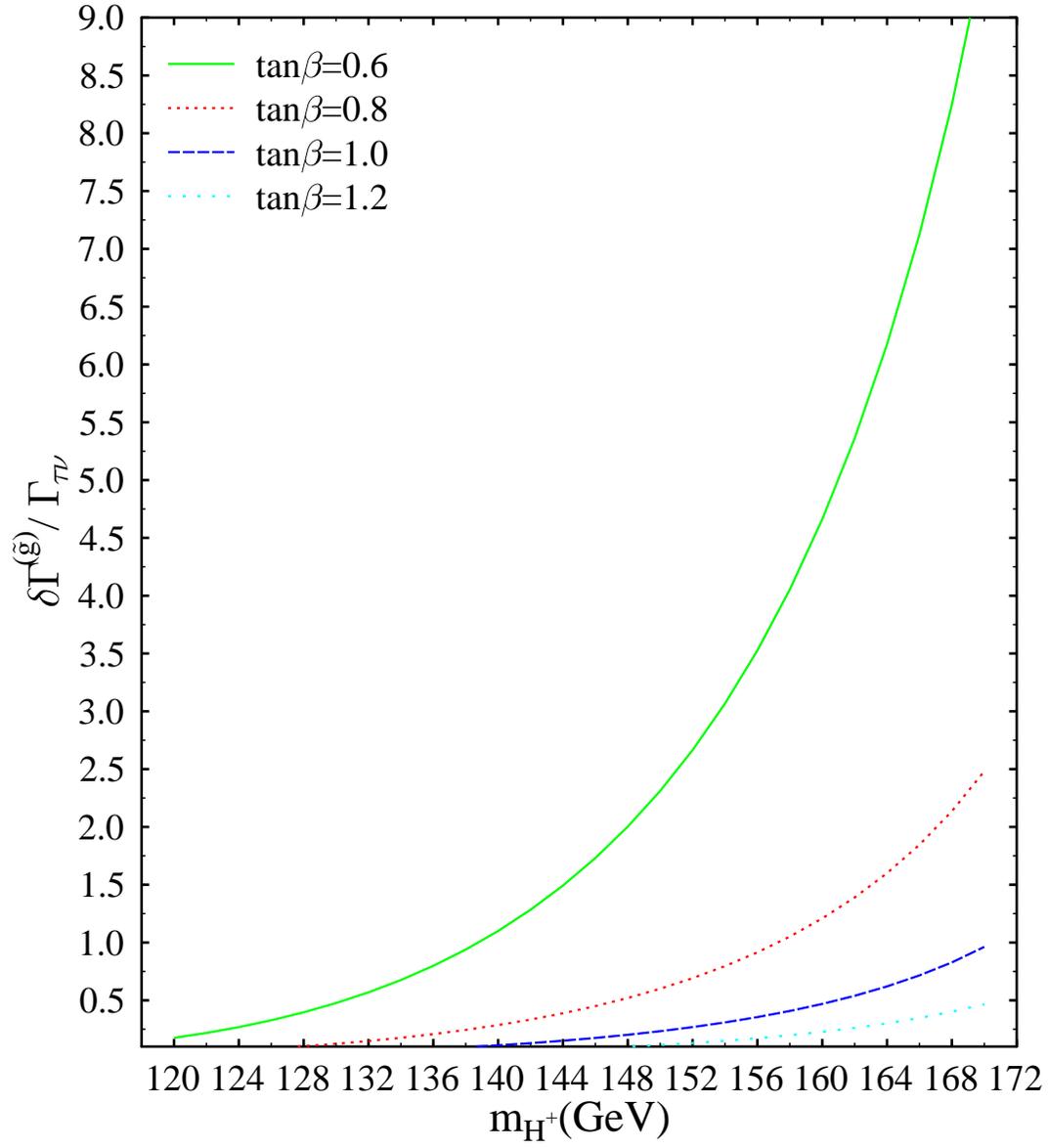
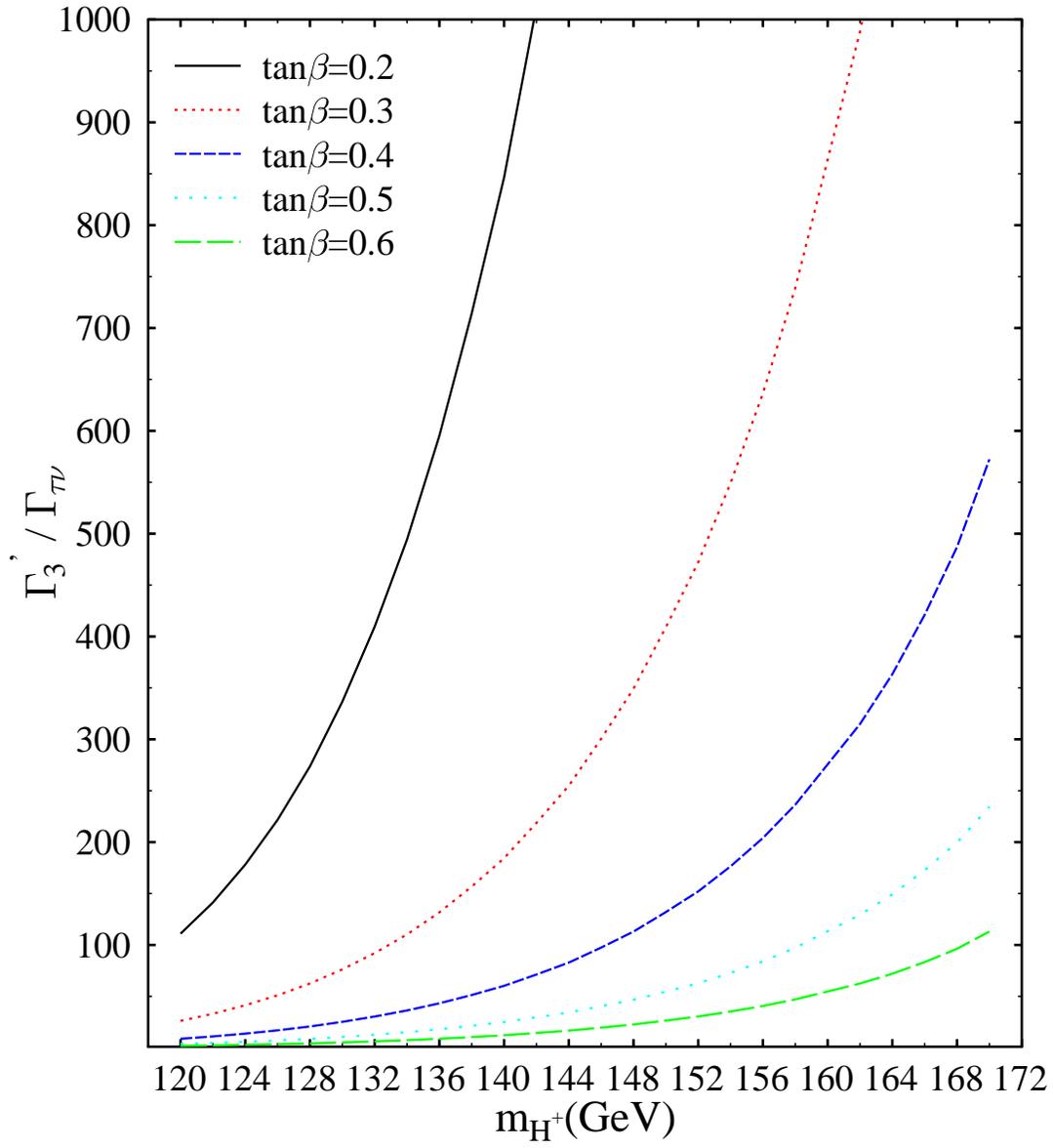
